# DASEL: Dark Sector Experiments at LCLS-II

Tor Raubenheimer, Anthony Beukers, Alan Fry, Carsten Hast, Thomas Markiewicz, Yuri Nosochkov, Nan Phinney, Philip Schuster, Natalia Toro

## ABSTRACT

This paper describes the concept for the DArk Sector Experiments at LCLS-II (DASEL) facility which provides a near-CW beam of multi-GeV electrons to the SLAC End Station A for experiments in particle physics. The low-current multi-GeV electron beam is produced parasitically by the superconducting RF linac for the LCLS-II X-ray Free Electron Laser, which is under construction at SLAC. DASEL is designed to host experiments to detect light dark matter such as the Light Dark Matter eXperiment (LDMX) but can be configured to support a wide range of other experiments requiring current ranging from pA to µA.

## 1 INTRODUCTION

The identity of dark matter (DM) is one of the most pressing open questions in fundamental physics today. Many searches have focused on Weakly Interacting Massive Particles (WIMPs) or extremely light DM particles such as axions. However, the possibility that dark matter particles have a mass similar to familiar matter, in the MeV-GeV range, is a largely unexplored region of parameter space. For an important class of light dark matter scenarios, electron fixed-target experiments are robust and have unparalleled sensitivity [1].

SLAC is constructing the LCLS-II X-ray Free Electron Laser (FEL) [2,3] for the photon science program. The LCLS-II is based on a 4 GeV CW superconducting RF linear accelerator and this presents a unique, timely, and cost-effective opportunity to enable high-impact dark matter and dark force experiments.

The proposal for DArk Sector Experiments at LCLS-II (DASEL) is to deliver a low-current, quasi-continuous electron beam into the SLAC End Station A (ESA) beamline by filling unused buckets from the LCLS-II linac. A new kicker and septum diverts the bunches from LCLS-II into a new transfer beamline and from there into ESA. Importantly, DASEL extracts beam downstream of the LCLS-II x-ray lines and, therefore, does not affect LCLS-II operations. Figure 1 is a photograph of the SLAC site showing the location of End Station A and the LCLS/LCLS-II enclosures.

DASEL's multi-GeV energy, high beam repetition rate, and capability to host year-scale particle physics experiments offer a unique combination of advantages that enable a wide range of world-class experiments. The first phase of DASEL supports sub-nA beam currents for the Light Dark Matter eXperiment (LDMX) [4,5,6]. LDMX is designed to decisively test thermal dark matter in the MeV-GeV mass range, a goal that no other existing or planned experiment can achieve [7]. The same beam may be useful for nuclear measurements relevant to the neutrino program and as a high-repetition-rate test beam. Upgrades to DASEL could support future experiments ranging from Super-HPS [8,9], a continuation of the current HPS experiment at JLAB, to a BDX-like beam dump experiment [10].

The following sections provide a technical overview of the facility, a description of possible experiments that would utilize DASEL, and then details of the DASEL design and layout of the facility.

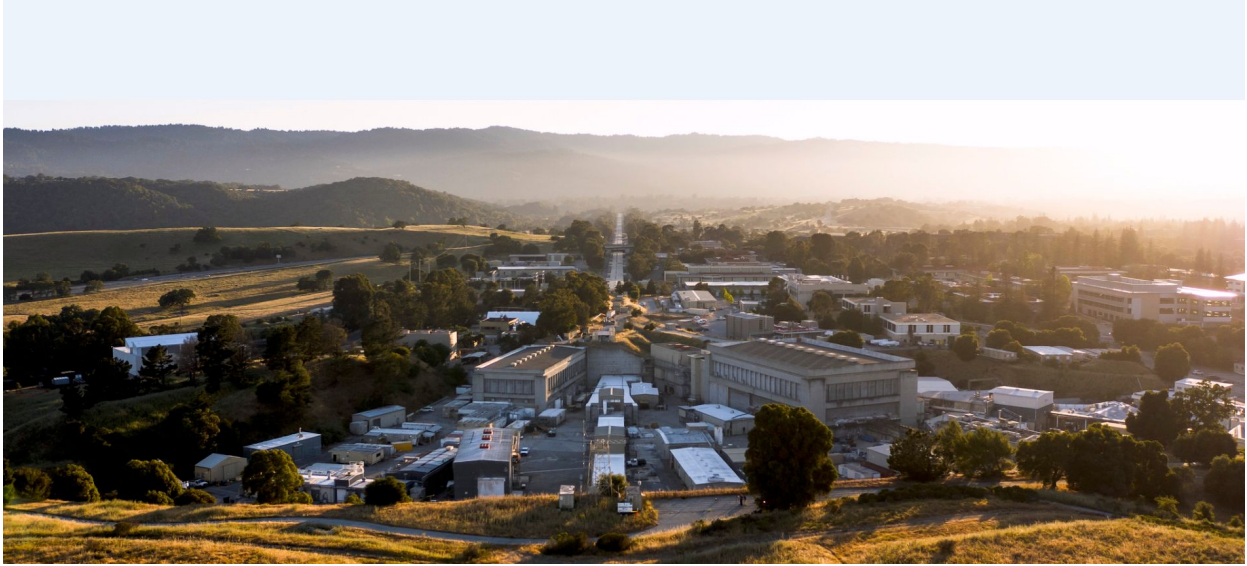

**Figure 1. Photograph looking upbeam and showing Research Yard with the SLAC linac gallery in the distance, the LCLS / LCLS-II enclosure following the straight linac line across the Research Yard, and End Station A (the large building to the right of the linac line).**

## 2  TECHNICAL OVERVIEW

This section summarizes the DASEL concept. DASEL uses the SLAC LCLS-II linac to provide a low-current, quasi-continuous beam to experiments in End Station A. The LCLS-II is an x-ray free electron laser based on a 4.0 GeV superconducting linac [2]. The linac operates with an RF frequency of 1.3 GHz and is fed from an RF gun [11] operating at up to 186 MHz, the seventh sub-harmonic of the RF linac. The baseline LCLS-II design has a maximum bunch rate of 929 kHz, corresponding to a bunch separation of 1,400 1.3-GHz RF buckets. Two high-speed kickers can deflect FEL bunches towards either the soft x-ray (SXR) or hard x-ray (HXR) undulators; unused beam travels to a high-power dump in the Beam Switch Yard (BSY). In initial operation, the LCLS-II linac accelerates up to 250 kW (nominally 62 µA at 4.0 GeV) of electrons to the BSY; an upgrade to the RF system can increase the beam current to 300 µA and the power to 1,200 kW. An energy upgrade to 8 GeV is also foreseen.

DASEL takes advantage of the "empty" RF buckets between LCLS-II bunches. These RF buckets are populated by a 46-MHz laser oscillator to produce a well-defined, low-current beam with 21.6 ns bunch spacing. The DASEL bunches are diverted to the DASEL beamline and sent to End Station A (ESA) with a third (new) kicker. A new 250-meter long beamline takes the bunches from the DASEL kicker/septum system to the existing ESA beamline, where the beam is further collimated. The secondary gun laser and a spoiler/collimation system control the charge delivered by DASEL. This is parasitic to LCLS-II operation, since the DASEL beam is low-current (<1µA compared to 62 µA nominal LCLS-II current), and is extracted downstream of the kickers that direct the primary beams to the undulators. The layout of the DASEL extraction is shown in Figure 2; the extraction concept is illustrated in Figure 3.

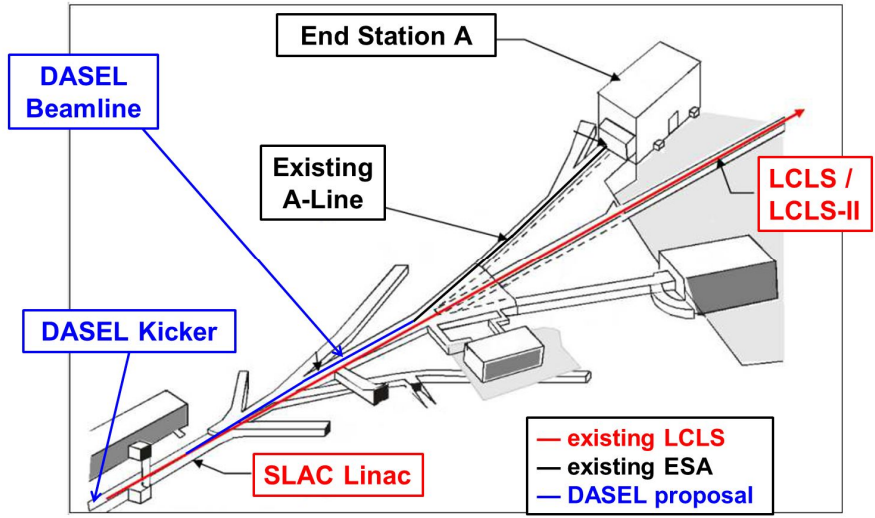

**Figure 2.** Layout illustrating SLAC linac, the LCLS / LCLS-II beamline, and End Station A with the DASEL extraction from the LCLS-II beamline.

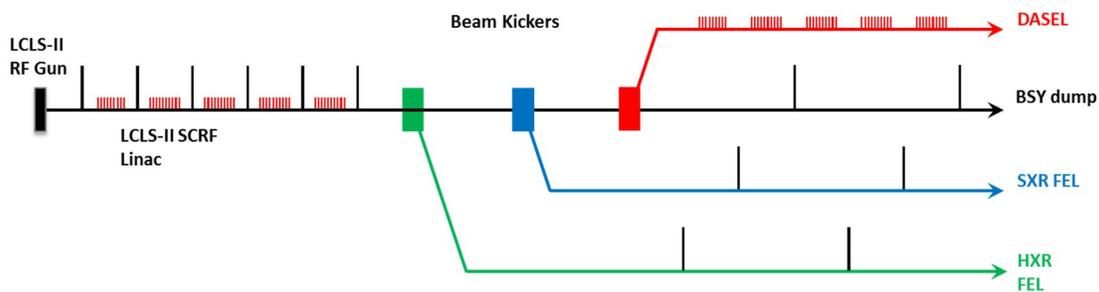

**Figure 3.** Schematic of the DASEL beam from the LCLS-II superconducting linac. The DASEL beamline directs unused beam to End Station A downstream of the extractions to the LCLS-II undulators.

## 3  SCIENCE CASE AND APPLICATIONS

End Station A and DASEL can support a wide array of experiments. Some examples are described here.

### 3.1  LDMX

The nature and origin of dark matter are among the foremost open questions in fundamental science today. Dark matter in the vicinity of Standard Model scales is simultaneously motivated by the viability of a thermal origin for its abundance and the existence of known matter at these scales. These motivations have led to a renewed interest over the last decade in searching for dark matter below GeV mass-scales (see e.g. [7,12,13]), where current direct detection experiments and LHC searches are not sensitive.

The flagship experiment envisioned for DASEL is the Light Dark Matter eXperiment (LDMX), which will search for sub-GeV dark matter by the "missing momentum" technique. The first phase of LDMX has world-leading sensitivity to light dark matter, and the second phase can fully explore the parameter space motivated by a thermal origin for dark matter. LDMX calls for a sub-nA, CW multi-GeV electron beam and is therefore very well-matched to DASEL capabilities. LDMX requires a beam spot spread over ~10 cm². This large spot size can be achieved by using the spoiler and collimator system in the A-line, as is done routinely for the End Station Test Beam program. The beam impinges directly on a silicon strip tracker and target inside a dipole analyzing magnet, which are used to measure the electron energy and transverse momentum before and after scattering. A downstream electromagnetic calorimeter and surrounding hadron calorimeter provide an inclusive energy measurement and hadron veto.

### 3.2 Nuclear Structure & Neutrino Physics

Accelerator-based neutrino experiments infer neutrino oscillation parameters from the energy spectrum of detected neutrinos; the neutrino's energy is determined from the energy of its scattering products, and this inference depends on nuclear physics modeling. Recent theoretical studies have demonstrated that nuclear physics modeling contributes significantly to the systematic uncertainties for multi-GeV neutrino experiments (see e.g. [14,15]), motivating new electron-beam measurements such as [16]. Studies are underway to understand the relevant uncertainties for DUNE. With a 4 GeV beam energy comparable to that of DUNE neutrinos, DASEL is well-suited to provide such measurements — for some final states, a detector modeled on LDMX, with modified target, may be especially complementary to low-acceptance measurements at other facilities like JLab.

### 3.3 High-Repetition-Rate Test Beam

The low-current ``LDMX-style'' DASEL beam also presents an opportunity for a new, high-repetition-rate test beam. This would enhance the capabilities of the existing End Station Test Beam facility [17] which typically supports ten to fifteen test beam experiments a year; some recent publications are listed in [18,19,20,21, and 22]. Many modern collider detectors (for example, at the LHC) operate at ns-scale repetition rates, presenting challenges for out-of-time pile-up. These must generally be addressed in situ, because test beam facilities have much lower event rates. A DASEL beam at 46 MHz (or a harmonic thereof) would enable test beam studies of detector performance in a high-repetition-rate environment.

### 3.4 Super-HPS

With an upgrade to ~1μA currents and 186 MHz repetition rate, DASEL would also enable searches for new GeV-scale force carriers (e.g. dark photons). The ``Super-HPS'' concept calls for a ~1 μA beam on a thin target; two spectrometers (similar to those used in the Heavy Photon Search) are placed downstream of a dipole magnet, enabling a high-statistics search for dark photons decaying promptly to leptons. This search significantly extends the parameter space for dark photons, potentially closing the parameter region between resonance- and vertex-based searches for dark photons [8,9].

### 3.5 Beam Dump Experiments

A third possible mode for DASEL running is motivated by recent interest in beam dump experiments to search for light dark matter or late-decaying force carriers [1,7,10,23,24,25,26,27]. Some of these experiments have complementary sensitivity to LDMX, while others could also follow up on an LDMX discovery. These beam dump experiments call for detectors downstream of a high power multi-GeV electron beam dump. This could be achieved at DASEL by diverting unused LCLS-II bunches from the BSY dump line to the DASEL line, then dumping it upstream of the End Station and placing experiments in the End Station. This configuration remains parasitic, and has two unique features: the large space and infrastructure available in End Station A, and the MHz spacing of bunches, which would facilitate using detector timing for time-of-flight measurements and/or rejection of cosmic backgrounds.

## 4 DASEL Design

DASEL uses the LCLS-II linac in a parasitic mode. The parameters of the DASEL system are listed in Table 1. Three categories of parameters are listed: beam at the experiment, beam in the End Station A beamline at the spoiler/collimator system, and beam in the LCLS-II accelerator. The primary beamline is designed to meet the LDMX experimental requirements with an ultra-low current beam but has the capability of being upgraded to support future higher-current Super-HPS type experiments and/or beam-dump experiments. Table 1 lists parameters for the ultra-low current beam and then for each of the potential upgrade paths. The ultra-low current beam has several applications beyond the LDMX experiment, including nuclear structure measurements motivated by the accelerator-based neutrino physics program and test beams.

For operation with ultra-low or low-current parameters, the DASEL kicker extracts roughly 600 ns of bunches between the LCLS-II primary bunches spaced by 1.1 µs, as illustrated in Figure 4. In the case of the LDMX experiment, the desired electron current ranges between 100 fA and 150 pA, corresponding to from 1 to approximately 500 electrons per µs, or a maximum of 0.5 Watts of electron beam power at 4 GeV with a 55% duty cycle. For the case of a low-current beam to support a Super-HPS type experiment, the beam current would be increased to an average of 1 µA which is still less than 2% of the nominal maximum current in LCLS-II. In this case, the spoiler would not be used, and the maximum beam power would be less than 5 kW into End Station A. Finally, in the case of a beam dump experiment, the DASEL kicker timing would be shifted to extract the primary bunches that are not sent to the FEL undulators. In operation, the photon science experiments are expected to use roughly ½ of the total beam power [2]; the excess electron bunches would be deflected into the A-line by the DASEL kicker and dumped upstream of End Station A. This upgrade would require installing a new 250 kW dump, adding the appropriate shielding, and plugging the aperture passing from the A-line through the 6m shield wall into the End Station A enclosure.

Focusing on the ultra-low current parameters, the buckets to be extracted by the DASEL kicker are filled at the RF gun. The LCLS-II RF gun specification [11] is that dark current is less than 400 nA at 100 MeV. All of the DASEL parameter sets stay below this current limit to exclude interference with LCLS-II. To ensure the performance required for LDMX and to enable higher currents in DASEL (as for Super-HPS), a separate gun laser is used. This intentionally populates unused gun buckets at a sub-harmonic of the gun frequency. These bunches are well-separated from the primary beam bunches so they can be extracted by the DASEL kicker downstream. The new gun laser shares the LCLS-II RF gun 46 MHz laser oscillator, but has a separate amplifier, UV conversion, and transport, all of which operate at much lower average laser power than the LCLS-II systems [28].

The desired beam emittance for the LDMX experiment is large enough so that the beam can be defocused to a cross-section of roughly 4x4 cm. The desired beam emittance is many times (100-1000) that of the LCLS-II emittance (as well as the LCLS-II admittance, which is determined by the collimation system). This increase is accomplished using the ESA spoilers with a corresponding degradation of the beam current. A spoiler system increases the beam emittance and the beam energy spread. Assuming an incoming, mono-energetic, 4-GeV beam, a simple 0.1 radiation length spoiler system increases the emittance to more than 300 µm, with more than 50% of the current within 0.5% of the incoming energy. In practice, DASEL has spoilers with different thicknesses which, combined with the downstream collimators, are used to control the beam emittance and current at the LDMX detector. The spoiler system has been specified for beam current up to 100 times higher than that needed at the experiment (i.e., 55 W) to allow options for precise control and shaping of the electron beam at the experiment.

**Table 1. DASEL electron beam parameters for an ultra-low-current beam (baseline) as well as two possible upgrade modes motivated by Super-HPS-style experiments and Beam-Dump experiments.**

| Experiment Parameters | Ultra-low-current | Low current (upgrade) | Dump-Style (upgrade) |
|---|---|---|---|
| Energy | 4.0 GeV (possible upgrade to 8.0 GeV) | 4.0 GeV (possible upgrade to 8.0 GeV) | 4.0 GeV (possible upgrade to 8.0 GeV) |
| Bunch spacing | 21.5 ns | 5.4 ns | LCLS-II nominal |
| Bunch charge | 0.04 – 20 e$^-$ | 70,000 e$^-$ (10 fC) | Up to 300 pC |
| Macro pulse beam current | 0.1 – 150 pA | 2 µA | Up to 62 µA |
| Duty cycle | 55% (600 ns out of 1.1 µs) | 55% (600 ns out of 1.1 µs) | Roughly 50%, depending on photon science experiments |
| Beam norm. emittance (rms) | ~100 µm; < 1000 µm | ~1 µm | <1 µm |
| Bunch energy spread | <1% | <1% | <1% |
| IP spot size | 4 cm x 4 cm | <250 µm including jitter | TBD |
| Max beam power | 0.5 W | 5 kW | 250 kW |
| | | | |
| **ESA Spoiler Parameters** | | | |
| Charge reduction | 0 – 99.99% | N/A | N/A |
| Emittance increase | 1 - 1000x | N/A | N/A |
| Max beam power | 55 W | N/A | N/A |
| Spoiler thickness | 0 – 0.5 r.l. | N/A | N/A |
| | | | |
| **Accelerator Parameters** | | | |
| Macro pulse beam current | 0 – 25 nA | 2 µA | N/A |
| Average current of diverted FEL bunches | N/A | N/A | Up to 62 µA |
| Beam norm. emittance (rms) | ~1µm; < 25 µm | ~1µm; < 25 µm | <1µm |
| Beam admittance (edge) | <50 nm, defined by LCLS-II collimators | <50 nm; defined by LCLS-II collimators | <50 nm; defined by LCLS-II collimators |
| Bunch energy spread (FWHM) | <2 % | <2 % | <2 % |
| Bunch length (rms) | <1 cm | <1 cm | <100 µm |
| Max beam power | 55 W | 5 kW | 250 kW |
| | | | Note: Exp. parameters in this column refer to beam on dump upstream of ESA |

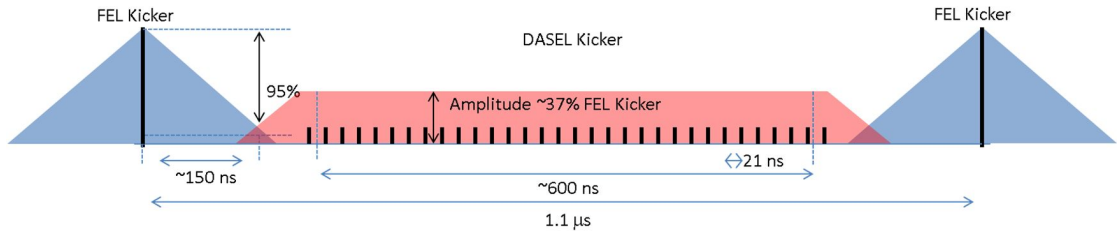

**Figure 4. LCLS-II pulse structure showing primary pulses with $4 \times 10^8$ e⁻ and DASEL bunches from the gun with ~30 e⁻ per bunch. The DASEL beam requires control of the bunch population with an additional seed laser and a spoiler/collimation system to deliver final current in the pA to µA range.**

# 5  DASEL INJECTOR LASER SYSTEM

DASEL uses the LCLS-II RF Gun which is based on the APEX gun developed at LBNL [11]. The RF Gun operates at 186 MHz and uses an excited CsTe photocathode. The DASEL laser system is co-located with the LCLS-II laser systems in a laser room upstream of Sector 0, approximately 20 meters from the photoinjector. The LCLS-II laser system [28] is based on a commercial laser architecture whose first element is a 46.4 MHz master oscillator producing 1040 nm pulses. For LCLS-II, pulses are selected at the nominal 0.929 MHz maximum repetition rate of the FEL, then amplified and shaped before a fourth harmonic generator (FHG) converts them to 260 nm.

DASEL calls for significantly lower power per pulse than LCLS-II, and has much looser requirements on pulse-to-pulse uniformity and pulse shape. In its first phase, DASEL uses a secondary output port from the 46.4 MHz LCLS-II oscillator (two such secondary ports are available). Several options for transporting the DASEL laser beam are under study, distinguished by where the DASEL amplifier, FHG, and acoustic-optic modulator (AOM) pulse picker are located.

A conceptual layout of the DASEL laser system is shown in Figure 5. The fiber delivering DASEL IR pulses arrives in the gun vault where the commercial "GOJI" amplifier, FHG and AOM units are located. Appropriate mirrors, half-wave plates and polarizers pick off diagnostic beams to power meters, photodiodes and a CCD camera before the beam is coupled collinearly with the LCLS-II beam for transport through the RF gun south-side window. Options that keep the DASEL active elements in the laser room and that share LCLS-II UV transmission to the gun vault are also being investigated. If subsequent experiments require electron pulses at the RF gun period of 5.4ns, a 186 MHz fiber laser oscillator and amplifier will have to be developed.

The full laser system was prototyped by the commercial supplier of the LCLS-II RF gun laser system. Starting from 2W of IR, the expected output of the laser oscillator, the laser was transported through a 60-m fiber and then recompressed to a pulse length of 280 fs FWHM. The beam was then focused to a 120 µm spot size at a type-I 4mm long Lithium Triborate crystal followed by a 3mm-long type-I Beta Barium Borate crystal to generate the 4th harmonic. Finally, a fast AOM pulse picker based on a TeO2 cell was implemented at the oscillator output to extract the desired 600 ns macro-pulse with a 55% duty cycle. The measured conversion efficiency at 2W of IR was > 2% as illustrated in Figure 6. The 40 mW of UV that was generated could produce an average current in excess of 10 µA, much higher than required for DASEL. The resulting spot profile is perfectly adequate for DASEL.

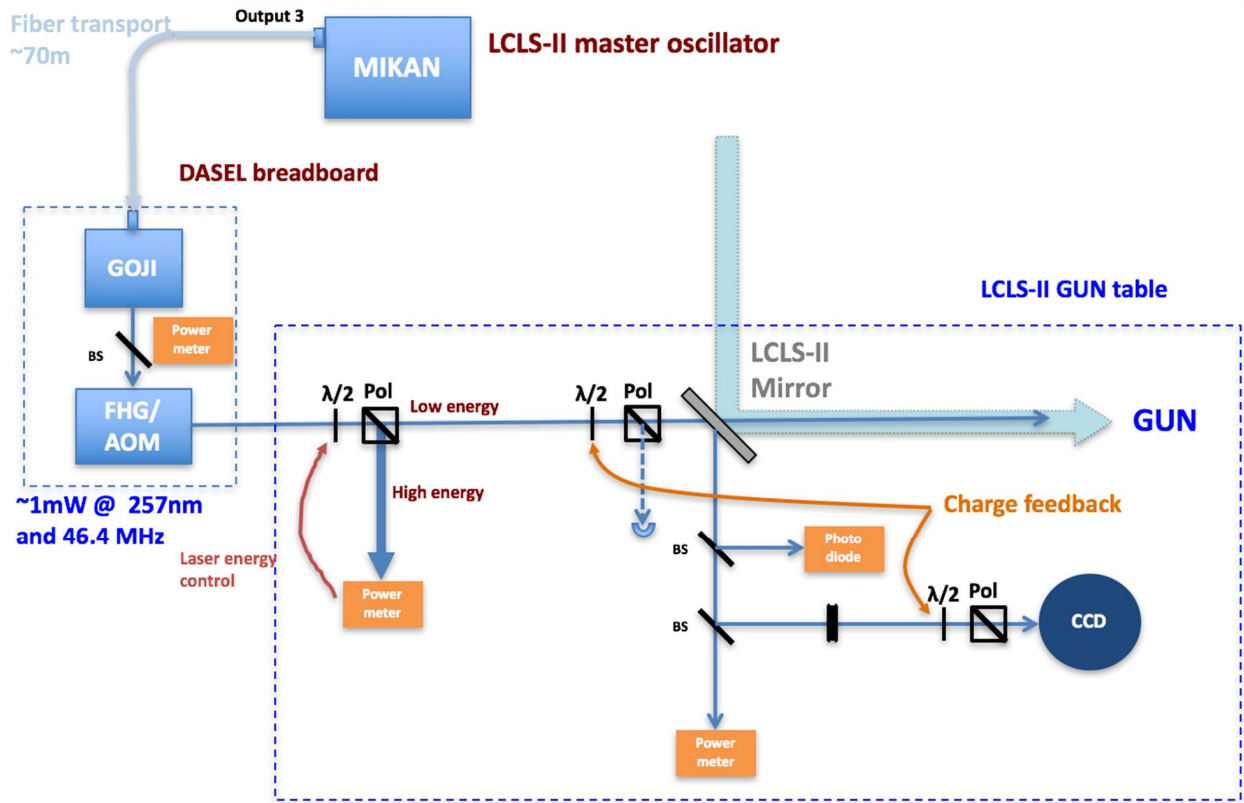

**Figure 5. A functional layout of the photoinjector region, showing the DASEL components required to produce 260nm pulses and DASEL diagnostics, before the DASEL laser beam couples co-linearly with the LCLS-II laser beam onto the cathode.**

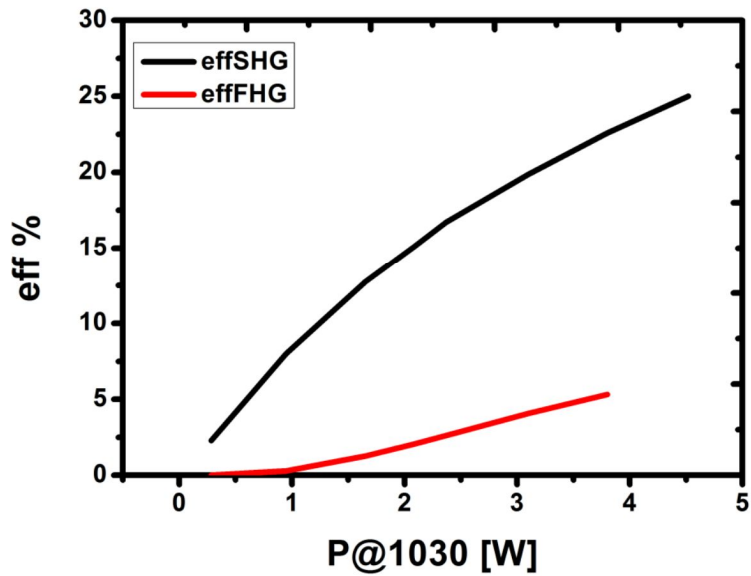

**Figure 6. Output efficiency at the 2nd and 4th harmonics as a function of the average input IR power in the demonstration by the commercial supplier of the LCLS-II RF gun laser system. With the expected input power of 2W, the efficiency was > 2% in the UV.**

# 6 DASEL Beamline

The DASEL beamline connects the BSY dump line to the existing A-line leading into End Station A (ESA) as illustrated in Figure 2. The detailed layout is complicated by other beamlines in the area and a plan view of the region is shown in Figure 7. This figure illustrates the layout of the LCLS-II beam spreader located between Sector-28 of the SLAC copper (CuRF) linac and the Beam Switch Yard muon shield wall; the location of the extraction lines to the HXR and SXR undulators; and the extraction lines to ESA for DASEL and the End Station Test Beam (ESTB). The spreader is a complicated region with several beamlines running through a 65x65-cm cross-section.

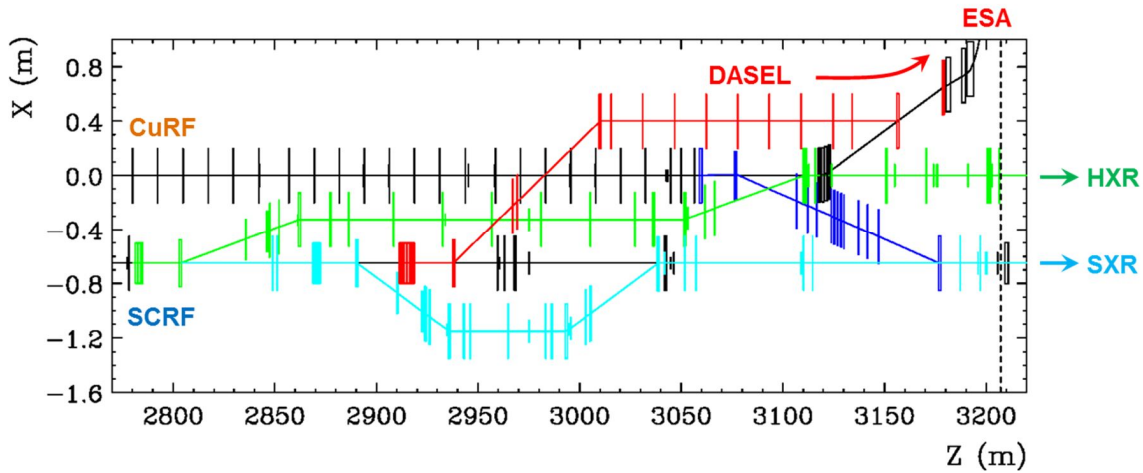

**Figure 7. Plan view of the LCLS-II spreader with the DASEL beamline. The beamline to the HXR undulator is green; the beamline to the SXR undulator is light blue (the link from the CuRF linac to the SXR is dark blue); the DASEL beamline is red; the elements in black belong to the CuRF linac, BSY dump line and A-line; the BSY dump is shown as a black box behind the BSY muon wall (vertical dashed line). Beamlines that appear to intersect are separated in elevation.**

To ensure that DASEL operation does not impact the LCLS-II FEL performance, the DASEL kicker is placed downstream of the HXR and SXR FEL kickers and septa. The kicker consists of six one-meter sections that direct the beam into a two-hole Lambertson septum magnet. A small vertical kick is used to send the beam to the septum hole with a strong horizontally deflecting field (for DASEL bunches); when the kick is off, the un-deflected main bunches pass through the field-free septum hole towards the BSY dump. After the septum, the DASEL line makes a horizontal cross-over above the HXR and CuRF linac beamlines, connecting to a DC-bend located 66.4 cm above and 40 cm to the left of the CuRF linac. This magnet is rolled, bending both horizontally and vertically, sending the beam downward at a shallow angle parallel to the CuRF linac. The arrangement of components in this region (about the 3010 meter location in Figure 7) is illustrated in Figure 8 which shows Top and Side views from the 3D CAD model. The beam line continues towards the second rolled DC-bend which merges the DASEL line with the A-line towards ESA (black line in Figure 7). The A-line is also connected to the HXR beamline for the transport of 120 Hz CuRF linac beam to ESA (ESTB project) using four pulsed magnets in the BSY. Fourteen quadrupoles in the DASEL beamline provide beam focusing and dispersion correction. The kicker induced orbit is compensated with a 2.6° roll of the kicker and septum magnets. A third rolled DC-bend in the beginning of the A-line is needed for compatibility with the ESTB. The DASEL optics functions are shown in Figure 9.

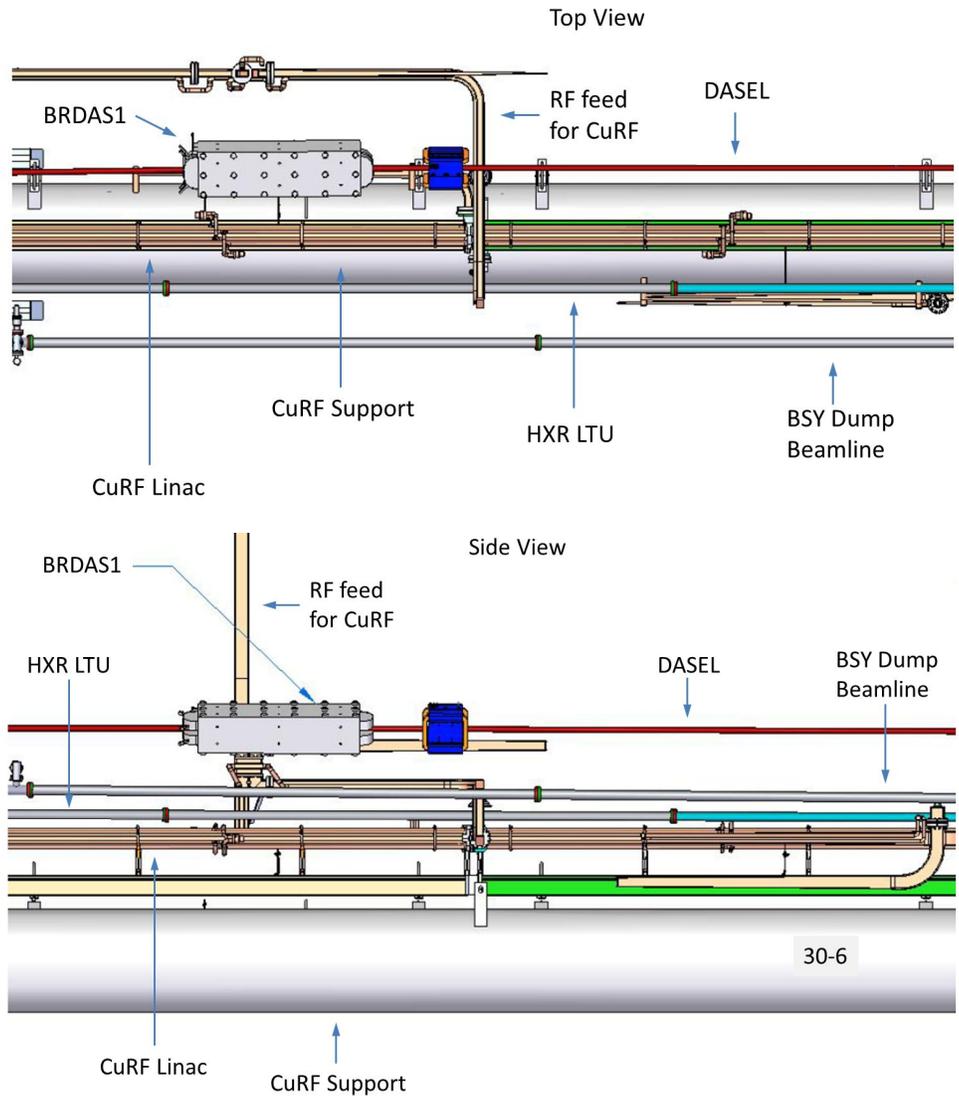

Figure 8. Top and Side views of the 3D CAD model in the region of the 1st rolled DASEL bend which is located around 3010 meters along the SLAC linac tunnel. Not shown is the SXR LTU beamline and the magnet support stands.

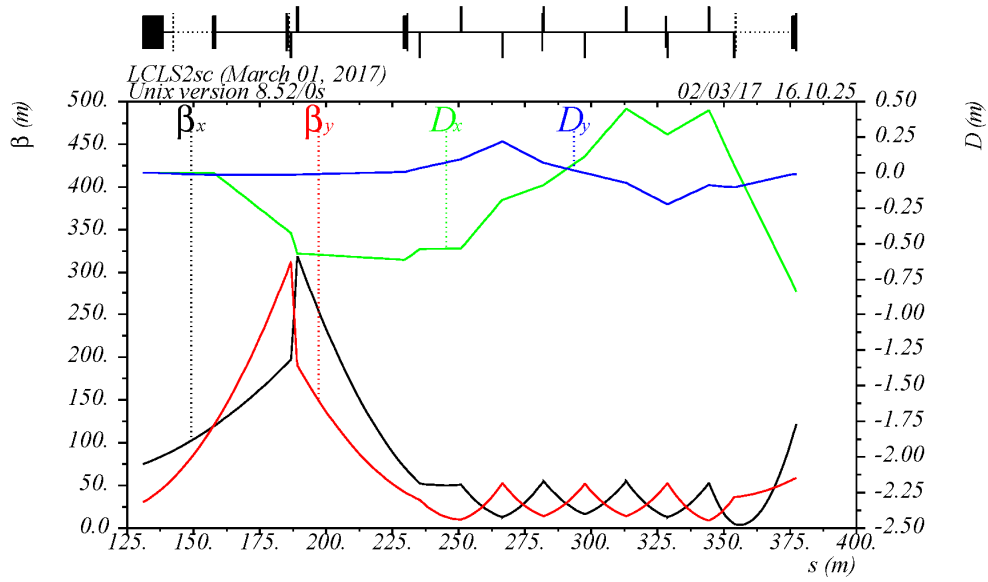

**Figure 9: Optics functions of the DASEL beamline.**

The DASEL beamline uses magnets already available at SLAC or existing magnet designs. These include the existing 1.0D38.37 and 2.0D38.37 dipoles, and the existing 2Q4W quadrupoles. The septum magnet is the same design developed for the LCLS-II HXR and SXR beamlines. The kicker design is based on the LCLS-II FEL kicker design. With a modest upgrade of the kicker, the DASEL magnets are compatible with 8 GeV beam energy. The magnet parameters are listed in Table 2. The 14 quadrupoles require nine independent power supplies; and the septum and three bend magnets need three power supplies.

**Table 2: Magnet parameters.**

|  | Quantity | Design | Aperture (mm) | Max. required field @ 8 GeV | Availability |
|---|---|---|---|---|---|
| Quadrupole | 14 | 2Q4W | 53.8 | 22.11 kG | Existing magnet |
| Kicker | 6 | 0.787K35.4 | 10 | 29.0 Gm per kicker | New, based on LCLS-II |
| Septum | 1 | 0.625SD38.98 | 15.9 | 3.89 kGm | New, based on LCLS-II |
| Rolled bend | 2 | 1.0D38.37 | 25.4 | 4.07 kGm | Existing magnet |
| Rolled bend | 1 | 2.0D38.37 | 50.8 | 0.62 kGm | Existing magnet |

The DASEL beam line diagnostic and correction system is shown in Figure 10. It includes one Beam Position Monitor (BPM) to control the kicker orbit, three profile monitors, six dipole correctors and two trims on two dipoles for orbit correction. Two additional BPMs are included as part of the Machine Protection System (MPS).

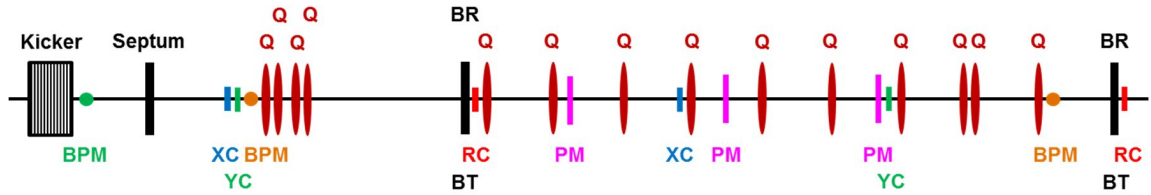

**Figure 10:** Schematic of the DASEL beamline with diagnostic and correction system. It shows the DASEL kicker; the Lambertson septum magnet; 2 rolled bends (BR); 14 quadrupoles (Q); a BPM for kicker orbit control (green); 2 MPS BPMs (brown); 2 horizontal (XC), two vertical (YC) and 2 rolled (RC) dipole correctors; 3 profile monitors (PM); and 2 bend trims (BT) on the rolled bends.

# 7 KICKER AND SEPTUM

The DASEL beam diversion system consists of a septum magnet and a vertical deflecting magnetic kicker. The kicker/septum combination sends beam towards the DASEL beamline. When the kicker is not energized, the beam traverses the zero field region of the septum magnet and is transported to the LCLS-II dump. The septum magnet is identical to the LCLS-II HXR and SXR Lambertson septum magnets.

The DASEL kicker operates at the same rate as the LCLS-II kickers but with a longer pulse, lower amplitude and looser tolerances. Allowing for the DASEL kicker rise/fall, roughly 600 ns of low current bunches can be extracted toward ESA between successive primary LCLS-II bunches spaced at 1.1 µs as illustrated in Figure 4.

The kicker is based on the design for the LCLS-II Spreader kicker. A ferrite loaded magnet topology reduces the voltage required to 1 kV. With this design, drive voltages can stay safely in the operating range of commercial MOSFETs. The magnet consists of serially connected L-C segments that approximate a transmission line. The inductive part of each section is made up of ferrites, gapped to accommodate the beam aperture. The capacitive part of each section is realized via discrete capacitors mounted in parallel on a printed circuit board. The capacitance can be easily adjusted to tune the magnet to the characteristic transmission line impedance of the system. Two copper bus bars provide conduction paths for supply and return current. Each magnet is composed of 18 segments. The DASEL kicker system includes six kicker magnets. Each one-meter magnet section contains an extruded ceramic metalized beam pipe to isolate the vacuum and provide uniform beam impedance. The prototype LCLS-II magnet with the ceramic chamber installed and the MOSFET-based pulser are shown in Figure 11. The resulting kicker pulse is shown in Figure 12. The DASEL kicker pulse length is extended by increasing the conduction time of the modulator MOSFETs. To compensate for the increased power dissipation, the kick strength of each section is reduced.

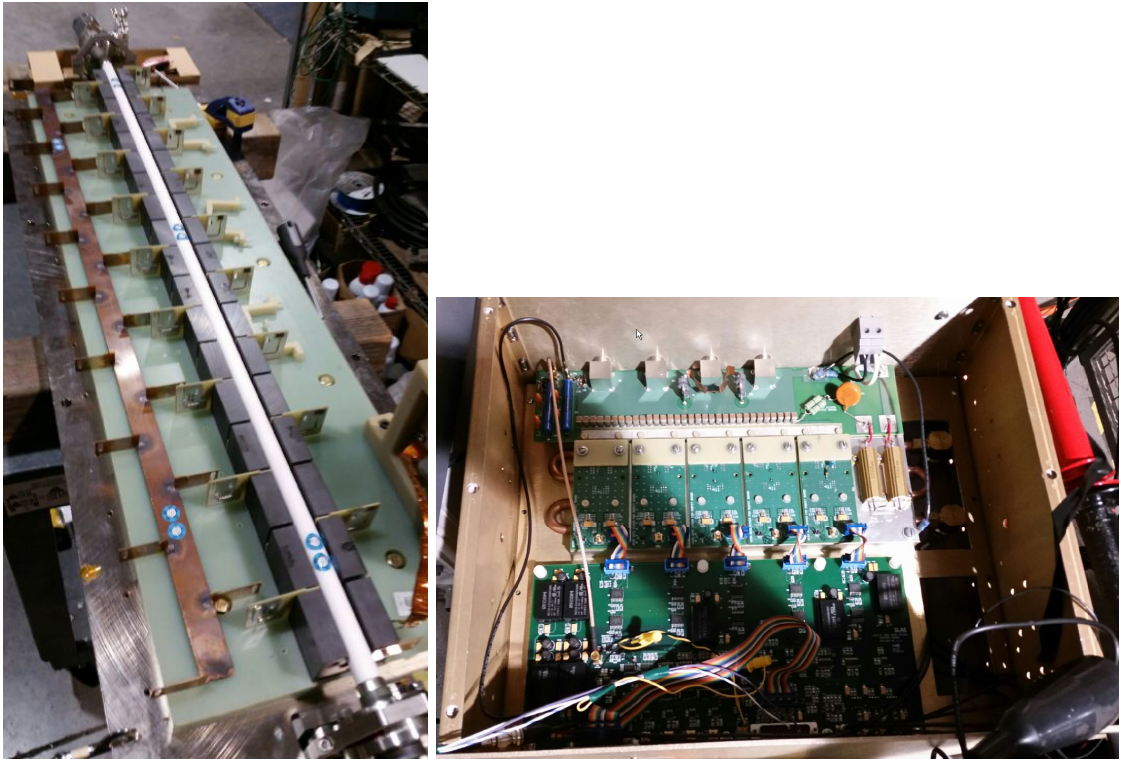

**Figure 11. Left: Prototype LCLS-II lumped-element kicker magnet with upper copper busbar removed and ceramic chamber installed. Right: MOSFET-based pulser to deliver 1 kV pulses.**

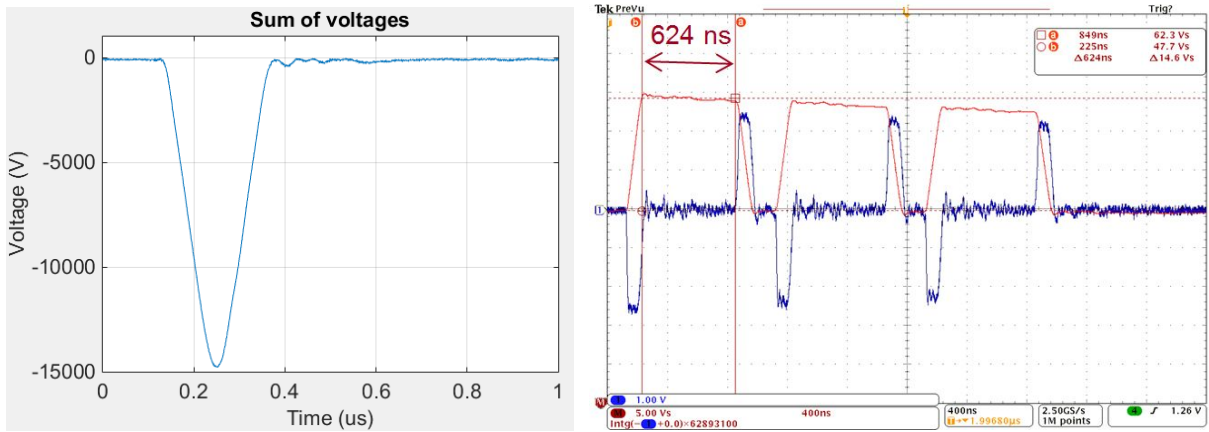

**Figure 12. Left: LCLS-II kicker pulse as measured on prototype kicker illustrating the total pulse width of roughly 250 ns. Right: a test demonstrating the increase an the kicker pulse width in excess of 600 ns by increasing the conduction time of the modulator MOSFETs; the visible droop across the kicker pulse will be corrected by increasing the storage capacitors.**

# 8 A-LINE AND END STATION A

The A-line, which transports beams from the central part of the Beam Switch Yard (BSY) towards End Station A (ESA), is presently set up to transport either the primary LCLS beam or a spoiled and collimated reduced number of electrons (secondary electron beam) from the LCLS beam into ESA. In 2013, the End Station Test Beam program (ESTB) [17] was established at SLAC to provide particle beams from the LCLS normal conducting Linac for detector R&D experiments in ESA. The A-Line ESTB has operated with the full range of available LCLS-beams: electron beam energies between 2.5 GeV and 16.5 GeV and bunch charges between 20 pC and 280 pC.

Presently, secondary electron beams are generated by steering selected LCLS bunches onto a 6-mm copper target in the BSY. The resulting electron beam has a wide energy spread, which is then transported through the A-line into ESA. The thickness and material of the target are chosen appropriately to reduce the number of hadrons generated in the electron-target interaction to negligible levels. Additional spoilers are available to further diffuse the electron energy spread without generating a significant number of other particles. The A-Line bend magnets are set to the particle energy required by the experimenters in ESA. Secondary particle beams have been delivered from 2 GeV up to the full-LCLS beam energy. A multi-collimator system in the A-line is used to control the number of electrons per pulse. A momentum slit reduces the accepted beam energy spread from about one percent to less than one part per million. Four-jaw collimators then reduce the geometrical spread of the accepted beam reaching End Station A.

DASEL operation takes advantage of the existing A-Line configuration as the beam delivery system for LDMX. Individual bunches impinge on a target. The resulting angle and momentum spread of the electrons allows operators to adjust the electron spot size and rate as required by the LDMX program.

## 9   DASEL Diagnostics and Tuning

The nominal DASEL bunch charge is too low to measure with standard LCLS-II diagnostics, such as Beam Position Monitors (BPM). Special electronics are used for the Average Current Monitors (ACMs) so that they can resolve the DASEL beam when the LCLS-II beam is turned off. The ACMs in the LCLS-II injector are used to set the DASEL laser amplitude.

Simulations show that the DASEL beam follows the LCLS-II beam trajectory closely. Thus special tuning through the linac beamline should not be necessary. The LCLS-II profile monitors are able to resolve the DASEL beam when the LCLS-II beam is off and can be used to confirm the trajectory and match. The LCLS-II wire scanners should also be able to resolve the DASEL beam when the LCLS-II beam is off and may be able to see it during LCLS-II operations by timing the detectors between the LCLS-II pulses.

The DASEL kicker and beamline tuning uses a special tune-up operating mode with low rate (1 ~ 10 Hz) of the LCLS-II primary beam. The primary beam can be extracted on pulses with the SXR and HXR undulator kickers turned off and the DASEL kicker timing shifted by 500 ns. In tune-up mode, the beam is stopped before it reaches the LDMX detector to prevent high charge bunches from damaging the detector electronics. To tune and align the beamline in this special configuration, three profile monitors and two BPMs are located along the length – see Figure 10. These are placed to confirm the dispersion, phase advance, and overall optical matching of the DASEL beam before it enters the ESA beamline.

During normal operation, the profile monitors can still be used to measure the DASEL beam. The two BPMs will not see the DASEL beam but are used to provide a very fast MPS signal to halt the DASEL kicker in case errant high charge bunches are extracted unexpectedly. In addition, the DASEL beamline has beam loss detectors along the length of the beamline and two ACMs. The ACMs are part of the Beam Containment System (BCS) and Machine Protection System (MPS) but are also used to measure the DASEL current upstream of ESA. After the initial setup described above, the DASEL current is maintained using this signal to feedback to the source laser. The loss monitors are used to provide a slow feedback signal for the kicker system, resolving slow drifts of the kicker amplitude.

DASEL commissioning would begin after the LCLS-II stably transports beam to the BSY Dump past the DASEL kicker. At that point, LCLS-II bunches are extracted to tune the DASEL beamline and ESA. After the beamlines are configured, the DASEL laser and ESA spoiler and collimator system are set up. The LCLS-II BCS Average Current Monitors (ACMs) have the resolution to detect 10 nA of current between the primary LCLS-II bunches at 929 kHz or 100 pA, when averaging without the LCLS-II beam. The initial setup of the laser is performed without the LCLS-II beam to establish 10~25 nA of current. The injector ACM is then used to maintain this level as the LCLS-II beam is re-established. The DASEL ACMs are used to verify the current extracted by the kicker; then the ESA spoiler and collimator system is configured. The final tuning of the ESA spoiler and collimation system is based on signals from the LMDX detector.

## 10 SUMMARY

For an important class of light dark matter scenarios, electron fixed-target experiments have unparalleled sensitivity. The proposal for DArk Sector Experiments at LCLS-II (DASEL) presents a unique, timely, and cost-effective opportunity to enable high-impact dark matter and dark force experiments. DASEL can deliver a low-current, quasi-continuous electron beam into the existing End Station A (ESA) beamline by filling unused buckets from the LCLS-II linac, without impacting the LCLS-II program. DASEL's multi-GeV energy, high beam repetition rate and capability to host year-scale particle physics experiments offer a unique combination of advantages that make possible a wide range of world-class experiments.

## 11 ACKNOWLEDGEMENTS

The authors would like to thank John Jaros for helping make the connection on the DASEL concept, the LDMX collaboration which has motivated the need for the facility, and the engineers and designers at SLAC that have developed the design. This work has been supported in part by DOE contract DE-AC02-76-SF00515.